\documentstyle[12pt,aasms4,psfig,epsfig]{article}

\newcommand{\lCCCHH}     {{\it l}-C$_{3}$H$_{2}$}
\newcommand{\cCCCHH}     {{\it c}-C$_{3}$H$_{2}$}
\newcommand{\ccchh}     {C$_{3}$H$_{2}$}
\newcommand{\lCCCH}     {{\it l}-C$_{3}$H}
\newcommand{\cCCCH}     {{\it c}-C$_{3}$H}
\newcommand{\ccch}     {C$_{3}$H}
\received{** 2000} \accepted{15/12/2000} 

\begin{document}

\def\th{$^{13}$}
\def\ei{$^{18}$}
\def\tw{$^{12}$}
\def\Lcs{{\hbox {$L_{\rm CS}$}}}
\def\lunits{K\thinspace \kms\thinspace pc$^2$}

\def\,{\thinspace}
\def\etal{et al.}


\def\kms{km\thinspace s$^{-1}$}
\def\Lsun{L$_\odot$}
\def\Msun{M$_\odot$}
\def\ms{m\thinspace s$^{-1}$}
\def\percc{cm$^{-3}$}

\font\sc=cmr10

\def\CBR{{\rm\sc CBR}}
\def\FWHM{{\rm\sc FWHM}}
\def\HI{{\hbox {H\,{\sc I}}}}
\def\HII{{\hbox {H\,{\sc II}}}}


\def\Ha{H$\alpha$}                      
\def\Htwo{{\hbox {H$_2$}}$\;$}
\def\nHtwo{n(\Htwo)}
\def\He#1{$^#1$He}                      
\def\water{H$_2$O$\;$}
\def\flecha{\rightarrow}
\def\COJ#1#2{{\hbox {CO($J\!\!=\!#1\!\rightarrow\!#2$)}}}
\def\CO#1#2{{\hbox {CO($#1\!\rightarrow\!#2$)}}}
\def\CeiO#1#2{{\hbox {C$^{18}$O($#1\!\rightarrow\!#2$)}}}
\def\CSJ#1#2{{\hbox {CS($J\!\!=\!#1\!\rightarrow\!#2$)}}}
\def\CS#1#2{{\hbox {CS($#1\!\rightarrow\!#2$)}}}
\def\HCNJ#1#2{{\hbox {HCN($J\!\!=\!#1\!\rightarrow\!#2$)}}}
\def\HCN#1#2{{\hbox {HCN($#1\!\rightarrow\!#2$)}}}
\def\HNC#1#2{{\hbox {HNC($#1\!\rightarrow\!#2$)}}}
\def\HNCO#1#2{{\hbox {HNCO($#1\!\rightarrow\!#2$)}}}
\def\HCOpJ#1#2{{\hbox {HCO$^+$($J\!\!=\!#1\!\rightarrow\!#2$)}}}
\def\HCOp#1#2{{\hbox {HCO$^+$($#1\!\rightarrow\!#2$)}}}
\def\HCOpp{{\hbox {HCO$^+$}}}
\def\J#1#2{{\hbox {$J\!\!=\!#1\rightarrow\!#2$}}}
\def\noJ#1#2{{\hbox {$#1\!\rightarrow\!#2$}}}


\def\Lco{{\hbox {$L_{\rm CO}$}}}
\def\Lhcn{{\hbox {$L_{\rm HCN}$}}}
\def\Lfir{{\hbox {$L_{\rm FIR}$}}}
\def\Ico{{\hbox {$I_{\rm CO}$}}}
\def\Sco{{\hbox {$S_{\rm CO}$}}}
\def\Ihcn{{\hbox {$I_{\rm HCN}$}}}


\def\Tastar{{\hbox {$T^*_a$}}}
\def\Tmb{{\hbox {$T_{\rm mb}$}}}
\def\Tb{{\hbox {$T_{\rm b}$}}}
\def\Tex{{\hbox {$T_{\rm ex}$}}}

\title{Molecular Carbon Chains and Rings in TMC-1}

\author {David Foss{\'e}$^{1,2}$, Jos{\'e} Cernicharo$^1$, Maryvonne Gerin$^2$, 
and Pierre Cox$^3$} 

\affil {$^1$ Depto F\'{\i}sica Molecular, I.E.M., C.S.I.C.,
Serrano 121, E-28006 Madrid, Spain}
\affil {$^2$ L.R.A., Observatoire de Paris \& Ecole Normale Sup{\'e}rieure, 24
rue Lhomond, F-75231 Paris Cedex 05, France}
\affil {$^3$ I.A.S., Universit{\'e} Paris-Sud, b{\^a}timent 121,
F-91405 Orsay, France}
\altaffiltext{1}{e-mail: David.Fosse@lra.ens.fr}

\begin{abstract}

We present mapping results in several rotational transitions of
HC$_{3}$N, C$_{6}$H, both cyclic and linear C$_{3}$H$_{2}$ and C$_{3}$H, 
towards the cyanopolyyne peak of the filamentary dense cloud
TMC-1 using the
IRAM 30m and MPIfR 100m telescopes. 
The spatial distribution of the cumulene carbon chain
propadienylidene H$_2$CCC (hereafter \lCCCHH\,) is found to deviate
significantly from the distributions of 
the cyclic isomer {\it c}-C$_{3}$H$_{2}$, HC$_{3}$N, and C$_{6}$H which in
turn look very
similar. The cyclic over linear abundance ratio of C$_3$H$_2$ 
increases by a factor of 3 across the filament, with a value of 28
at the cyanopolyyne peak.
This abundance ratio is an order of magnitude larger 
than the range (3 to 5)
we observed in the diffuse interstellar medium.
The cyclic over linear 
abundance ratio of C$_3$H also varies by $\sim$ 2.5 in
TMC-1, reaching a maximum value (13) close to the cyanopolyyne peak.
These behaviors
might be related to 
competitive processes between ion-neutral and
neutral-neutral reactions for cyclic and linear species.  
\\

\end{abstract}

\section{Introduction}

Among the molecules discovered in the interstellar medium, C$_{3}$H$_{2}$ and C$_{3}$H
are of peculiar interest for astrochemistry since both are observed in two isomeric forms: cyclic and 
linear.
The ring molecule cyclopropenylidene (hereafter \cCCCHH\,) was discovered in 1985
(Thaddeus, Vrtilek, \& Gottlieb) and has 
immediatly retained attention due to its
ubiquity in the galaxy (e.g., Matthews and Irvine 1985). One of its linear 
counterparts, the propadienylidene (\lCCCHH\,), was discovered in TMC-1 
by Cernicharo
\etal\,(1991). The two isomeric shapes of C$_{3}$H, cyclopropynylidyne 
(\cCCCH\,) and propynylidyne (\lCCCH\,), were also detected in TMC-1 
(Yamamoto \etal\, 1987; Thaddeus \etal\, 1985).

This 
variety of isomers for the same species raises the question of their 
formation. In particular, are rings 
and chains formed from the same progenitors and involved in the 
same reactions networks? Furthermore  
these isomers
can be used to probe interstellar chemistry
models which include heavy interstellar molecules such as cumulene carbenes 
(Bettens and
Herbst 1996, 1997; see also Millar, Leung, and Herbst 1987). 
Previous observations by Cernicharo \etal\, (1999) have shown that
the cyclic over linear
abundance ratio of C$_{3}$H$_{2}$ (hereafter R$_2$) in the diffuse medium 
along the line of sight toward
the continuum sources W51E1/E2, W51D and W49 is one order of magnitude
smaller than its value in TMC-1. In a recent study, Turner, Herbst, \& Terzieva (2000)
have compared measurements in three translucent clouds
and two dense clouds (TMC-1 and L183) of the C$_3$H cyclic over linear abundance 
ratio (R$_1$) and R$_2$
ratio - and column densities of 6 other
hydrocarbons - to a modified version of the New Standard Model of chemistry
(Lee \etal\ 1996). They found small variation of R$_1$ from source to source, and a
systematic higher R$_2$ ratio with largest the values found in the two dark clouds. 
Also, it has been suggested that 
cyclic and linear isomers of a same compound could have different behavior in 
neutral-neutral and ion-neutral reactions (see, for example, Adams \& Smith 1987;
Maluendes, McLean, \& Herbst 1993; Kaiser \etal\ 1997, 1999). 
The cyclic over linear abundance ratio of molecules like C$_{3}$H$_{2}$ and 
C$_{3}$H could then be used as a tool to investigate physical conditions in several
media, from cold dark cloud to warmer and lower density media.

In order to investigate these questions, we mapped in detail the region
around the cyanopolyyne peak (hereafter CP) in TMC-1 in both 
\lCCCHH\, and \cCCCHH, and observed at selected positions the two 
isomeric forms of C$_3$H. C$_{6}$H and HC$_3$N have also been observed for
comparison purpose.

\section{Observations}

The C$_3$H$_2$ and C$_6$H observations were made in 1990 and 1992 with
the 100-meter 
telescope of the Max-Planck Institut f{\"u}r Radioastronomie at 
Effelsberg (Germany). We observed the $\rm 1_{10}-1_{01}$ transitions of 
\cCCCHH\ 
at 18343.145 MHz, and simultaneously 
those of $\rm C_6H \,  ^2\Pi_{3/2}$ $\rm J \, = \, 15/2-13/2$ near 20794 MHz
and \lCCCHH\ $\rm 1_{01}-0_{00}$ at 20792.590 MHz.
 The antenna beamsize at 18.3 and 20.7 GHz are of 
$\rm 54^{\prime\prime}$ and $\rm 48^{\prime\prime}$, respectively. We used the 1024
channel autocorrelator to achieve 
a resolution of 0.05 $\rm km \, s^{-1}$ at 18.3 GHz and 0.09 $\rm km \, s^{-1}$ at 
20.7 GHz, and
the data were taken by position 
switching the telescope. 

The HC$_3$N and C$_3$H observations were done in 1995 and 1999 with the 
IRAM 30-meter telescope at Pico Veleta 
(Spain). We observed the \lCCCH\, $\rm ^2\Pi_{3/2}$ $\rm J \, = \, 9/2-7/2$ 
transition near 98 GHz and at the same time the transition of \cCCCH\ 
$\rm 2_{12}-1_{11}$ near 91.5 GHz.
The observations were made in frequency-switching mode. The autocorrelator 
was used as a spectral
instrument with a velocity resolutions of 0.12 kms$^{-1}$.
The half-power beamwidth (HPBW) and main-beam efficiency are 26" and 0.75 for 
\lCCCH\ and 27" and 0.78 for \cCCCH. System 
temperatures were 
in the range 90-130K. For HC$_3$N, observed at 90979.023 MHz ($\rm J \, = \, 10-9$), the
HPBW is 26". Pointing and calibration were monitored 
by regularly
observing planets and quasars for both telescopes.
Sample spectra are shown in Fig.1 and line parameters for selected positions are given in
Table 1.

\section{Results}
\subsection{C$_3$H$_2$ isomers and C$_6$H}

The mapping results of this study are shown in
Fig.~2. The distributions towards the CP in TMC-1
of \cCCCHH\, and \lCCCHH\, are compared to that of
C$_6$H in Fig.~2a and b, respectively. Fig.~2c 
shows the distributions of C$_6$H and HC$_3$N. The 
spatial  distributions of \lCCCHH\, and \cCCCHH\, (Fig. 2d) are 
clearly different, whereas the emissions of
\cCCCHH, C$_6$H and HC$_3$N have comparable distributions.
The emission of \lCCCHH\, is shifted toward the west by 
approximatively 40'' with 
respect to \cCCCHH\, or C$_6$H - note
that, since 
C$_6$H and \lCCCHH\, were observed in the same bandwidth, this shift  
cannot be due to pointing errors during the observations. 

In order to derive relative abundances, we first computed relations between 
the column-density and the observed line intensity valid for uniform physical conditions,
i.e. n(H$_2$) = 3$\times $10$^4$ cm$^-$$^3$ and T$_K$ = 10K in TMC-1 (Cernicharo \& 
Gu{\'e}lin 1987). 
From a statistical equilibrium
calculation using collisional excitation rates for \cCCCHH\ from Avery \& Green 
(1989), 
and assuming an ortho/para  
ratio of 3, we 
find:
 {\it N}(\cCCCHH)[cm$^{-2}$] = 2.2$\times $10$^{13}$ $\int T_{mb}dv$[K$\cdot$km$\cdot$s$^{-1}$] . 
This linear relationship indicates 
that the $\rm 1_{10}-1_{01}$
transition of \cCCCHH\ is optically thin in TMC-1; it is in good
agreement with the work of Cox \etal\ (1989). Madden \etal\ (1986), 
from observations of 
the isotopic {\it c}-C$_2$$^{13}$CH$_2$ in TMC-1, 
derive an optical depth 
ranging from 5.6 to 6.8 for the $\rm 2_{12}-1_{01}$ transition of \cCCCHH\ at 85 GHz. 
This result 
is in agreement with ours since the $\rm 2_{12}-1_{01}$ line has a larger opacity than the
$\rm 1_{10}-1_{01}$ transition for dark cloud physical conditions. Indeed, with our code, we
are able to reproduce the results of Madden \etal\ at 85 GHz, the $\rm 1_{10}-1_{01}$ transition 
still being 
thin or marginally saturated
($\tau$ $\leq$ 1.5). 
For \lCCCHH\ and C$_6$H there are 
no collisional cross sections available. We have estimated cross sections from those of H$_2$CO 
(calculated by Green et al. 1978) and of HC$_3$N (Green \& Chapman 1978) - 
see Cernicharo et al. (1999).
This gives: {\it N}(\lCCCHH)[cm$^{-2}$] =
2.2$\times $10$^{13}$ $\int T_{mb}dv$[K$\cdot$km$\cdot$s$^{-1}$] and {\it N}(C$_6$H)[cm$^{-2}$] = 
5.2$\times $10$^{13}$ $\int T_{mb}dv$[K$\cdot$km$\cdot$s$^{-1}$] .
Note that for C$_6$H, $ \int T_{mb}dv$ is the 
total 
integrated intensity summed over
the four components of the hyperfine structure. These relationships are in good 
agreement with previous works (Cernicharo \etal\, 1991;
Bell \etal\, 1999).

Comparing two positions separated by 40'' (which corresponds to 0.02 pc for the 
adopted distance of 100 pc to TMC-1 - Cernicharo \& Gu{\'e}lin 1987), we derive 
the following column density: at the CP (0,0),  
{\it N}(C$_6$H) = 8.3$\times $10$^{12}$ cm$^{-2}$, 
{\it N}(\cCCCHH) = 5.8$\times $10$^{13}$ cm$^{-2}$, 
{\it N}(\lCCCHH) = 2.1$\times $10$^{12}$ cm$^{-2}$ (R$_2$=28$\pm $6); and 
at the edge of the TMC-1 filament (--40,0),
{\it N}(C$_6$H) = 4.7$\times $10$^{12}$ cm$^{-2}$,
{\it N}(\cCCCHH) = 2.8$\times $10$^{13}$ cm$^{-2}$,
{\it N}(\lCCCHH) = 3.2$\times $10$^{12}$ cm$^{-2}$ (R$_2$=10$\pm $3). Between
these two positions, the cyclic over linear
abundance ratio changes by a factor of $\sim$ 3.
This variation cannot be an artefact caused by 
calibration errors. Indeed, while the
calculated C$_6$H over {\it l}-C$_3$H$_2$ 
column density ratio varies from 1.5 to 4, the {\it c}-C$_3$H$_2$ over C$_6$H column density 
ratio remains constant ($\simeq$ 6.5).
Since the former ratio is unaffected by
calibration and pointing errors (lines are observed in the same bandwidth), the observed variation 
of R$_2$ must be real. This 
result suggests that chemical gradients are present in TMC-1 on scales smaller than
0.02 pc.

To investigate further the spatial variations of molecular abundances across the TMC-1 
filament, 
we averaged the spectra along the six different rows shown in Fig.2. 
Results are summarized in Fig.3: R$_2$ ranges from 12 to 37 and clearly increases from the west to the east
of TMC-1. Note that a constant density of
n(H$_2$) = 3$\times $10$^4$ cm$^{-3}$ has been adopted. 
If we use instead H$_2$ densities derived by 
Pratap \etal\ (1997) 
from an analysis of the 
HC$_3$N transitions while keeping the same kinetic temperature (10 K), the values of R$_2$ are then 
lowered by 20 to 40 \% but the west to east variation of R$_2$ remains. The 
rise of the cyclic 
over 
linear
abundance ratio of the C$_3$H$_2$ isomers at a spatial scale of 2 arcmin (0.06 pc) is
therefore a firm result.

\subsection{C$_3$H isomers}

As C$_3$H is thought to be formed by the same reaction as C$_3$H$_2$, i.e.
by the dissociative recombination of C$_3$H$_{3}^{+}$ 
(Adams \& Smith 1987),
it is interesting to 
compare {\it N}(\cCCCH)/{\it N}(\lCCCH) (hereafter R$_1$) with R$_2$. What is
the value of R$_1$ and does it vary across the filament similarly to R$_2$?
In order to answer these questions we observed {\it c}-C$_3$H and
{\it l}-C$_3$H at 13 positions along two strips close to the CP.
One strip crosses the filament from offset positions (60",60") to 
(--60",--60"), the
other from (10,--70") to (--70",10"). We also observed the \cCCCHH\ 
and the \lCCCHH\ peaks.
Where possible, we analyzed the data with the HFS method of CLASS (a software developed by
the GILDAS working group). This method provides the total optical depth, the average 
linewidth and the brightness temperature of a line with hyperfine structure. A
 reliable estimate of the excitation temperature could be obtained for 
several points. We found that 3K $\leq$ T$_{ex}$ $\leq$ 3.8K for \cCCCH\ 
(in excellent agreement with 
Mangum \& Wootten 1990), and 4.9K $\leq$ T$_{ex}$ $\leq$ 6.7K
for \lCCCH. In the following, we adopt
\Tex = 3.5K for \cCCCH\ and \Tex = 5.5K for the linear isomer.  

Calculations of the column densities were done using the classical formula with
$\mu $ = 2.4 D (Yamamoto et al. 1987) for \cCCCH\ and
$\mu $ = 3.1 D (Green 1980) for \lCCCH. 
For each observed point, we computed the average of the total
column densities by using each hyperfine transition
weighted by the inverse of the square of the error. 
The resulting column densities and the values for R$_1$ are given in Table 2. 
Comparing with the values of R$_2$ in Fig.3, we see that: 1) With a value of
12 at the CP, R$_1$ is smaller than R$_2$; 2) As
R$_2$ does, R$_1$ shows variations across the filament (by a factor 
of 2.5), mainly due to variations of {\it N}(\cCCCH).

\section{Discussion}

The observations described above underscore the differences in the distribution 
and relative abundances of the cyclic and linear forms of \ccchh\, and \ccch\, in 
the molecular filament TMC-1. In the following, we study the origin of this
behavior and suggest that it is driven by competitive processes between 
ion-neutral and neutral-neutral reactions.

\subsection{Steady-state calculation}

We assume here, as it has been proposed (see, for example, Adams \& Smith 1987;
Maluendes, McLean \& Herbst 1993), that cyclic C$_3$H$_2$ and C$_3$H are
both formed in the dissociative recombination of c-C$_3$H$_{3}^{+}$, while linear 
C$_3$H$_2$ and C$_3$H result from H$_2$C$_3$H$^+$.
We also assume that the C$_3$H$^{+}$ + H$_2$ association reaction 
is the dominant formation mechanism of  
C$_3$H$_{3}^{+}$, which is produced in equal amounts in cyclic and linear forms.  

At steady state, the chemical kinetic equation for
the abundance of \cCCCHH\, - that is, x(\cCCCHH) 
- is:
\begin{center}
dx(\cCCCHH)/dt = $^2$b$_c$k$^{r}_{c}$x(c-C$_3$H$_{3}^{+}$)x$_e$ --
$^2$K$^{d}_{c}$x(c-C$_3$H$_{2}$) = 0 
\end{center}
where "b" is for "branching ratio", k$^{r}$
is for "recombination rate" [cm$^3$$\cdot$s$^{-1}$], K$^{d}$ is 
for "mean destruction rate" [s$^{-1}$] (K$^{d}$ = $\sum_i k^d_i \cdot x_i$ , where x$_i$ 
is the second reactant).
We use indices (l,c) and
exponants (2,1) to distinguish between C$_3$H$_2$ and C$_3$H in linear and
cyclic forms (for example, $^2$b$_c$ means "branching ratio
for {\bf c}yclic C$_3$H$_{\bf 2}$"). Similar equations can be written for \lCCCHH, \cCCCH\, and 
\lCCCH. Using the observed column densities for each species at the border 
of the
filament (i.e. row number 2), where the electron density should be 
the highest in our data set, and
writing k$^{r}_{c}$ = k$^{r}_{l}$ and X $\equiv$ 
x(c-C$_3$H$_{3}^{+}$)/x(l-C$_3$H$_{3}^{+}$),
calculations lead to: 
\begin{center}
x(\cCCCHH)/x(\cCCCH) = 
($^2$b$_c$/$^1$b$_c$)($^1$K$^{d}_{c}$/$^2$K$^{d}_{c}$) $\simeq$ 5 ;\\ 
x(\lCCCHH)/x(\lCCCH) = ($^2$b$_l$/$^1$b$_l$)($^1$K$^{d}_{l}$/$^2$K$^{d}_{l}$) 
$\simeq$ 2 ;\\
R$_2$ = ($^2$b$_c$/$^2$b$_l$)($^2$K$^{d}_{l}$/$^2$K$^{d}_{c}$)X $\simeq$ 16 ;\\
R$_1$ = ($^1$b$_c$/$^1$b$_l$)($^1$K$^{d}_{l}$/$^1$K$^{d}_{c}$)X $\simeq$ 6 . 
\end{center}
A simple and coherent set of solution for this system is 
$^2$b$_l$ $\simeq$ $^1$b$_l$ $\simeq$ $^1$b$_c$ 
$\simeq$ $^2$b$_c$ ; $^1$K$^{d}_{l}$ $\simeq$ 2 $^2$K$^{d}_{l}$ ; 
$^1$K$^{d}_{c}$ $\simeq$ 5 $^2$K$^{d}_{c}$ ; 
$^1$K$^{d}_{c}$ $\simeq$ (1/6) $^1$K$^{d}_{l}$ X ;
$^2$K$^{d}_{c}$ $\simeq$ (1/16) $^2$K$^{d}_{l}$ X .

Under the hypothesis of steady state and 
from the observed abundance ratio at the border of the 
TMC-1 filament, we find that \cCCCHH\ is destroyed approximatively
5 times slower than \cCCCH\ and that the \lCCCH\ mean destruction rate is twice 
the \lCCCHH\ one. 

To our knowledge, no additional formation process of \cCCCHH\ in dark clouds 
can be invoked to explain an
increase of its abundance with respect to \lCCCHH.
Variations of R$_2$ (and R$_1$) in TMC-1 could result from variations of the destruction
rates of \cCCCHH\ and \lCCCHH. Such an explanation has been proposed by Maluendes, McLean,
\& Herbst (1993): while \cCCCHH\ is inert with respect to most neutral-neutral
reactions, \lCCCHH,
as \lCCCH\, and \cCCCH, are easily destroyed by these reactions. Hence, because
neutral-neutral 
reactions proceed fast when a radical reacts with
abundant atoms, it is possible in a dark cloud like TMC-1 that R$_2$ is one
order of magnitude larger than in a more diffuse medium where the proportion of 
reactive ions (increasing with respect to reactive atoms) is sufficient to destroy 
\cCCCHH\ and \lCCCHH\ at the same rate.
The same explanation holds for the progenitor ions c-C$_3$H$_{3}^{+}$ and
l-C$_3$H$_{3}^{+}$ (see Cernicharo \etal\ 1999): 
while
l-C$_3$H$_{3}^{+}$ - assumed to be the progenitor of \lCCCHH\ - can be efficiently
removed through ion-neutral reactions lowering the amount available 
to produce \lCCCHH,
c-C$_3$H$_{3}^{+}$ is mainly affected by dissociative recombination 
to produce \cCCCHH.
R$_2$ thus depends on the ion-neutral reactions of l-C$_3$H$_{3}^{+}$ that
do not affect the cyclic ion: in dark clouds, where the abundance of 
reactive molecules
can be hundred times larger than that of electrons, l-C$_3$H$_{3}^{+}$ 
is removed faster 
by ion-neutral reactions than by the dissociative recombination lowering the production rate
of \lCCCHH, and increasing R$_2$. In a
more diffuse medium, where the ion-neutral reactions are much less efficient, 
l-C$_3$H$_{3}^{+}$ and
c-C$_3$H$_{3}^{+}$ are removed by the same process and R$_2$ is then closer to 
$^2$K$_{l}^{d}$/$^2$K$_{c}^{d}$ (whose value also decreases with decreasing 
density as suggested above). This explanation is strengthened by the 
lower R$_2$ values found in the diffuse medium from absorption measurements by 
Cernicharo \etal\ (1999) - 3 to 5 versus 10 to 40 in TMC-1 - and by the observed R$_1$ 
variations across the filament.
We test this hypothesis in the next section.

\subsection {Chemical modelling}

In order to study the impact of physical conditions on the R$_2$ ratio, we
have run 25 models of gas phase chemistry with different densities and visual
extinctions using a time dependent chemistry code solving the system of
stiff ordinary differential equations with the Gear method.
Although we have used a time dependent chemistry code - samples of the
evolution chemistry are shown in Fig.4 - we discuss here the results at steady
state. Indeed, we would like to study the basic processes of the isomeric
differentiation, and to identify the dominant reactions leading to the observed 
large variations of R$_2$ in diffuse and dense clouds. We aim first at a qualitative
description of the mechanism. A quantitative description 
would require a more accurate knowledge of the reaction rates (including branching 
ratios) important for \cCCCHH\ and \lCCCHH. Fig.4 shows that \cCCCHH/\lCCCHH\ and
c-C$_3$H$_{3}^{+}$/l-C$_3$H$_{3}^{+}$ abundance ratios reach values close to the
steady state value, just after they reach their maximum abundance. As a first step,
it is thus reasonable to consider only the processes
leading to variation in isomeric abundances at steady state.
We have used, as chemical network, the UMIST95 database (Millar, Farquhar, 
\& Willacy 1997) in a pure gas-phase scheme. As recommanded by the authors, the
species C$_2$H$_5$, C$_2$H$_6$ and their ions have been excluded from 
the network. We have also excluded the species including the
following atoms: P, Si, Cl, Na, Mg. On the other hand, we have included the grain  
surface formation of H$_2$ with a rate coefficient 1.5$\times $10$^{-17}$ n$^2$(H) 
cm$^{-3}$ s$^{-1}$. Moreover, we have updated the dissociative recombination rate of 
c-C$_3$H$_{3}^{+}$ at
300 K to 7.0$\pm $2.0 $\times $ 10$^{-7}$ cm$^3$ s$^{-1}$ (Abouelaziz \etal 1993) 
and assumed the same
value for l-C$_3$H$_{3}^{+}$. Parameters of the models and initial elemental 
abundances are given in
Table 3 and 4, respectively.

In Fig.4, we see that R$_2$, and also X, increase with increasing visual extinction. 
This general trend of our models agrees with observations since R$_2$ is lower in
the diffuse medium  
than in the TMC-1 dark cloud. The models also show that the increase is 
faster in higher density media.
We have plotted in Fig.5 the electronic density with respect to R$_2$ for each model. The
distribution of the points exhibit clearly a correlation between the ionization fraction 
and the cyclic over linear ratio which can be separated in two regimes: "low" 
(1$\leq$A$_V$$\leq$2) and "high" (5$\leq$A$_V$$\leq$10), {\it A$_V$} = 3 mag 
being an intermediate
case. In the low extinction regime, R$_2$ strongly depends on the electronic abundance. 
This is an interesting result because it opens the possibility to use R$_2$ as a tool
to probe the electronic abundance in low extinction regions where H$^{13}$CO$^+$ and DCO$^+$ 
cannot be detected.
Why is the cyclic over linear ratio sensitive to the ionization fractionation? 
In the UMIST95 chemistry, C$^+$ can destroy both \lCCCHH\ and \cCCCHH, whereas C reacts 
only with the linear isomer.
R$_2$ is then sensitive to the electronic abundance {\it via} the C$^+$/C ratio as shown 
in Fig.6. Note that the range 3 to 5 that Cernicharo \etal\ (1999) have observed in 
the diffuse medium 
corresponds, as expected, to high electronic abundances and to moderate visual extinctions. In
the high extinction regime, R$_2$ is independent of the C$^+$/C ratio, while it is sensitive 
to the (H$_3^+$+C$^+$)/O ratio (Fig.7). Indeed, in the UMIST95 database the atomic oxygen reacts
only with \lCCCHH\ (O + \lCCCHH\ $\rightarrow$ C$_2$H$_2$ + CO). This reaction is negligible
with respect to other destruction reactions (in particular the reactions with C$^+$ which 
affects both isomers) until all
the carbon is locked into CO. It is thus unefficient in low visual extinction media where 
the abundances of  C and  C$^+$ remain high. On the other hand, it becomes
the principal destruction channel of \lCCCHH\ at high visual extinctions: for 
A$_V$ = 
2 mag and a density of 10$^3$ cm$^{-3}$ the O + \lCCCHH\ reaction is roughly
15\% of the \lCCCHH\ destruction rate while it represents more than 80\% of it at
A$_V$ = 10 mag.
The variations of the cyclic over linear abundance ratio of C$_3$H$_2$ can 
thus be understood as a
consequence of the competition between neutral-neutral and ion-neutral reactions in the 
interstellar medium.

The determination of electronic abundances from measurements of R$_2$ using Fig.5 may not be
straightforward. Indeed, taking into account the large uncertainties 
in the reaction rates and branching ratios, it is difficult to infer quantitative results from
this study on the C$_3$H$_2$ isomers. 
We can say however that low R$_2$ values are specific of high electronic abundance 
while 
high R$_2$ values indicate low electronic abundance. In a more general way, isomeric ratio can be used to
probe physical conditions in different media from the diffuse gas to dark clouds.
In the future,
cyclic and linear species other than the isomers of C$_3$H$_2$ shall be included in chemical 
models. C$_3$H is a good candidate.
Indeed, Kaiser \etal\ (1997, 1999) have shown that the neutral-neutral reaction
C + C$_2$H$_2$ $\rightarrow$ C$_3$H + H produces linear and cyclic C$_3$H, with an increase 
of the linear over cyclic abundance ratio with rising collision energies. It would be 
interesting to make measurements of R$_1$ in absorption in the diffuse medium 
in order to compare this neutral-neutral
formation path with the C$_3$H$_{3}^{+}$ dissociative recombination.

This study has been partly begun. Indeed, in a recent paper, Turner, Herbst, \& 
Terzieva 
(2000) investigate the R$_1$ and R$_2$
ratios in three translucent clouds and in the dark clouds TMC-1 and L183. They 
compare observations to 
the predictions of a modified version of the New Standard Model of chemistry 
(Lee \etal 1996)
which includes cyclic and linear species for all precursor to C$_3$H and C$_3$H$_2$.
One of their conclusion is that the
clear observed anticorrelation of R$_2$ with extinction suggests that the ratios are
strongly affected by formation or destruction rates, and not just by branching ratios
among the relevant chemical reactions. 

\section{Conclusion}

In this paper, we have presented new results about the cyclic over linear abundance ratios of
C$_3$H$_2$ and C$_3$H toward the CP of TMC-1. These ratios exhibit variations 
which are probably due to variations in the electronic abundance across the filament. Indeed, the
chemical modelling of the reactions that affect \cCCCHH\ and \lCCCHH\ shows that their ratio 
is sensitive mainly to the behavior of the C$^+$/C (at low visual extinction) and 
(H$_3^+$+C$^+$)/O (at high visual extinction) abundance ratios, which are both related 
to the electronic 
abundance. As C$_3$H$_2$ is observed in its cyclic and linear form in different physical 
conditions,
it could be used as a tool to probe the electronic abundance in the ISM. However, to infer 
accurate values of the fractional ionization from cyclic and linear isomers abundances, 
the branching ratios and kinetic rates used in chemical networks must be more accurately known. 
In addition to laboratory studies, a systematic 
measurement of R$_1$ and R$_2$ in different astrophysical media should be useful 
for this work.

{\it Acknowledgements}
We thank C.M. Walmsley for useful comments and 
suggestions.
D.F. and J.C. thank the Spanish DGCYT for fundings under grants N$^{\circ}$PB96-0883 and PNIE98-1351E.
D.F. also acknownledges the Spanish Ministry of Foreign Affairs for support 
under grant N$^{\circ}$195.

\newpage

\begin{center}
\tiny{
\begin{tabular}{ccccccccccc}
\multicolumn{11}{c} {Table 1: Line Parameters for
C$_6$H, {\it l}-C$_3$H$_2$ and {\it c}-C$_3$H$_2$ at Selected Positions in TMC-1}\\
\hline
&C$_6$H&&&{\it l}-C$_3$H$_2$&&&&{\it c}-C$_3$H$_2$ \\
\cline{2-7}\cline{9-11}
Position & T$_{mb}$& v$_{LSR}$& $\Delta$v& T$_{mb}$& v$_{LSR}$& $\Delta$v&
Position & T$_{mb}$& v$_{LSR}$& $\Delta$v\\
( " , " ) & K & km s$^{-1}$& km s$^{-1}$ &  K & km s$^{-1}$& km s$^{-1}$ &
( " , " )&  K & km s$^{-1}$& km s$^{-1}$ \\
\hline
0,0&0.41$\pm $0.02&5.72$\pm $0.01&0.27$\pm $0.02&0.27$\pm $0.03&5.83$\pm $0.02&
0.33$\pm $0.03&0,0&3.69$\pm $0.59&5.54$\pm $0.01&0.12$\pm $0.02\\
&0.13$\pm $0.02&6.02$\pm $0.05&0.29$\pm $0.07&&&&&4.44$\pm $0.59&5.72$\pm $0.01&0.14$\pm $0.02\\
&&&&&&&&5.46$\pm $0.59&5.89$\pm $0.02&0.20$\pm $0.02\\
&&&&&&&&2.46$\pm $0.59&6.12$\pm $0.02&0.14$\pm $0.03\\
-40,0&0.31$\pm $0.03&5.74$\pm $0.01&0.25$\pm $0.00&0.31$\pm $0.07&5.78$\pm $0.03&0.44$\pm $0.08&
-50,0&2.20$\pm $0.76&5.48$\pm $0.05&0.20$\pm $0.00\\
&0.11$\pm $0.11&5.99$\pm $0.02&0.09$\pm $0.20&&&&&3.99$\pm $0.76&5.70$\pm $0.03&0.20$\pm $0.00\\
0,40&0.29$\pm $0.04&5.82$\pm $0.03&0.21$\pm $0.06&0.28$\pm $0.06&5.85$\pm $0.02&0.18$\pm $0.04&
0,50&3.96$\pm $0.71
&5.68$\pm $0.02&0.17$\pm $0.04\\
&0.25$\pm $0.04&6.07$\pm $0.04&0.21$\pm $0.09&0.26$\pm $0.06&6.15$\pm $0.02&0.14$\pm $0.04&&
4.58$\pm $0.71&5.96$\pm $0.02
&0.28$\pm $0.05\\
40,-40&0.48$\pm $0.03&5.78$\pm $0.01&0.25$\pm $0.00&0.28$\pm $0.07&5.80$\pm $0.03&0.21$\pm $0.06&
50,-50&
2.83$\pm $0.78&5.61$\pm $0.03&0.19$\pm $0.05\\
&0.22$\pm $0.03&6.03$\pm $0.03&0.27$\pm $0.05&0.23$\pm $0.07&6.08$\pm $0.03&0.13$\pm $0.06&&
4.88$\pm $0.78&5.88$\pm $0.03&
0.34$\pm $0.06\\
\hline
\multicolumn{11}{l} {\footnotesize{\tiny{Notes -- Positions are offsets 
relative to the Cyanopolyyne Peak : $\alpha$(1950)
= 04$^h$38$^m$38.6$^s$, $\delta$(1950) = 
25$^{\circ}$35'45.0''}}}\\
\multicolumn{11}{l} {\footnotesize{\tiny{TMC-1 exhibits two main components at
v$_{LSR}$ $\simeq$ 5.3 and 6.1 km s$^{-1}$ (Sume, Downes, \& Wilson 1975)}}}\\
\multicolumn{11}{l} {\footnotesize{\tiny{When possible, we have fitted several components along a line of sight}}}\\
\end{tabular}
}
\end{center}

\newpage

\begin{center}
\begin{tabular}{cccc}
\multicolumn{4}{c} {Table 2: Derived Column Densities}\\
\multicolumn{4}{c} {for \cCCCH\, and \lCCCH\, in TMC-1}\\
\hline
Position&{\it N}(\cCCCH)&{\it N}(\lCCCH)&{\it N}c/{\it N}l\\
( " , " )&10$^{12}$ cm$^{-2}$&10$^{11}$ cm$^{-2}$& (= R$_1$)\\
\hline
-40,-40&3.7$\pm $0.2&$\leq$4&$\geq$9\\
-20,-20&7.7$\pm $0.3&8.0$\pm $0.2&10$\pm $0.5\\
0,0&10.3$\pm $0.3&8.4$\pm $0.1&12$\pm $0.5\\
20,20&10.0$\pm $0.3&7.6$\pm $0.4&13$\pm $1.2\\
40,40&5.8$\pm $0.3&6.7$\pm $0.5&9$\pm $1.1\\
60,60&3.6$\pm $0.2&6.1$\pm $0.7&6$\pm $1.0\\
\hline
10,-70&5.2$\pm $0.2&7.0$\pm $0.4&7$\pm $0.6\\
-10,-50&5.3$\pm $0.2&7.8$\pm $0.2&7$\pm $0.4\\
-30,-30&5.1$\pm $0.2&8.0$\pm $0.1&6$\pm $0.4\\
-50,-10&4.7$\pm $0.2&7.1$\pm $0.5&7$\pm $0.8\\
-70,10&3.7$\pm $0.2&7.3$\pm $0.3&5$\pm $0.5\\
\hline
30,-60$^{ 1}$&7.8$\pm $0.3&8.0$\pm $0.2&10$\pm $0.6\\
-30,0$^{ 2}$&10.0$\pm $0.4&8.2$\pm $0.2&12$\pm $0.7\\
\hline
\multicolumn{4}{l} {\footnotesize{\small{Notes -- Positions are offsets relative to the CP :}}}\\
\multicolumn{4}{l} {\footnotesize{\small{$\alpha$(1950)
= 04$^h$38$^m$38.6$^s$, $\delta$(1950) =
25$^{\circ}$35'45.0''}}}\\
\multicolumn{4}{l} {\footnotesize{\small{$^{1}$ \cCCCHH\ peak}}}\\
\multicolumn{4}{l} {\footnotesize{\small{$^{2}$ \lCCCHH\ peak}}}\\
\end{tabular}
\end{center}

\newpage

\begin{center}
\begin{tabular}{lr}
\multicolumn{2}{c}{Table 3: Parameters of the Chemical Models}\\
\hline
Parameter&Value\\
\hline
n(H$_2$) (10$^3$ cm$^{-3}$)&1, 3, 10, 30, 100\\
T (K)&10\\
A$_V$ (mag)&1 , 2 , 3 , 5 , 10\\
$\zeta$ (s$^{-1})$&1.3 $\times $ 10$^{-17}$\\
{\it u$_i$} (erg cm$^{-3}$)& 2 $\times $ 10$^{-15}$\\
\hline
\multicolumn{2}{l} {\footnotesize{\small{$\zeta$ is the cosmic-ray ionization rate}}}\\
\multicolumn{2}{l} {\footnotesize{\small{{\it u$_i$} is the density of ionizing radiation 
at A$_V$ = 0}}}
\end{tabular}
\end{center}

\newpage

\begin{center}
\begin{tabular}{lr}
\multicolumn{2}{c}{Table 4: Initial Elemental Abundances}\\
\multicolumn{2}{c}{ with respect to Hydrogen}\\
\hline
He&1.4 $\times $ 10$^{-1}$\\
C&7.3 $\times $ 10$^{-5}$\\
N&2.14 $\times $ 10$^{-5}$\\
O&1.76 $\times $ 10$^{-4}$\\
S&2.0 $\times $ 10$^{-8}$\\
Fe&3.0 $\times $ 10$^{-9}$\\
\hline
\end{tabular}
\end{center}

\clearpage
\begin{center}
Figure Captions
\end{center}

\figcaption[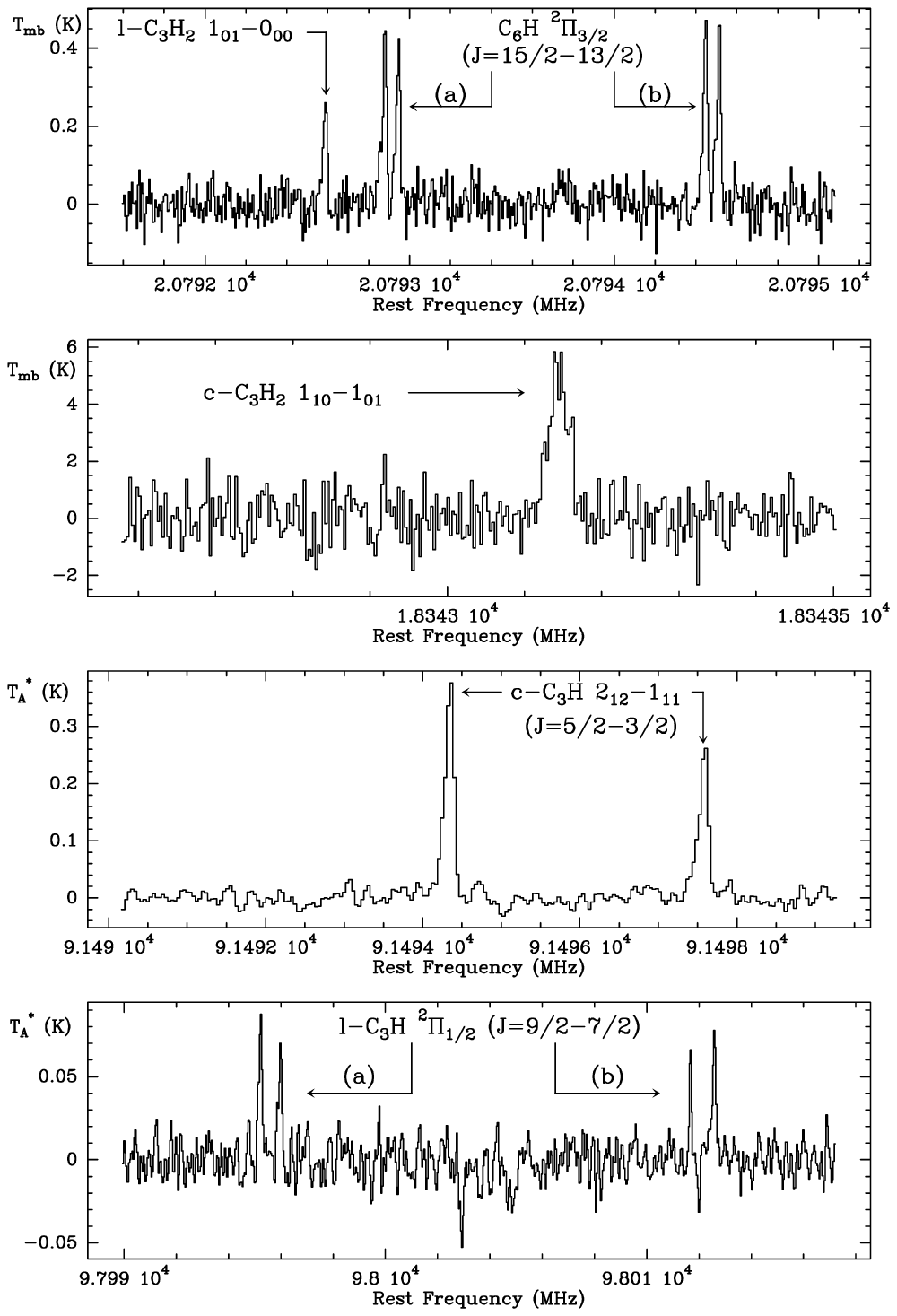]
{Spectra of C$_6$H, cyclic and linear C$_3$H$_2$, and cyclic and linear C$_3$H
towards the cyanopolyyne peak of TMC-1
(v$_{LSR}$ = 5.8 kms$^{-1}$). Note that
the C$_6$H $^2{\Pi}_{3/2}$
J=15/2--13/2 quadruplet
and the {\lCCCHH} 1$_{01}$-0$_{00}$ transition are observed in the same bandwidth.
The J=3/2--1/2 F=2--1 and F=1--0 transitions of {\cCCCH} 2$_{12}$--1$_{11}$
are not shown here.}

\figcaption[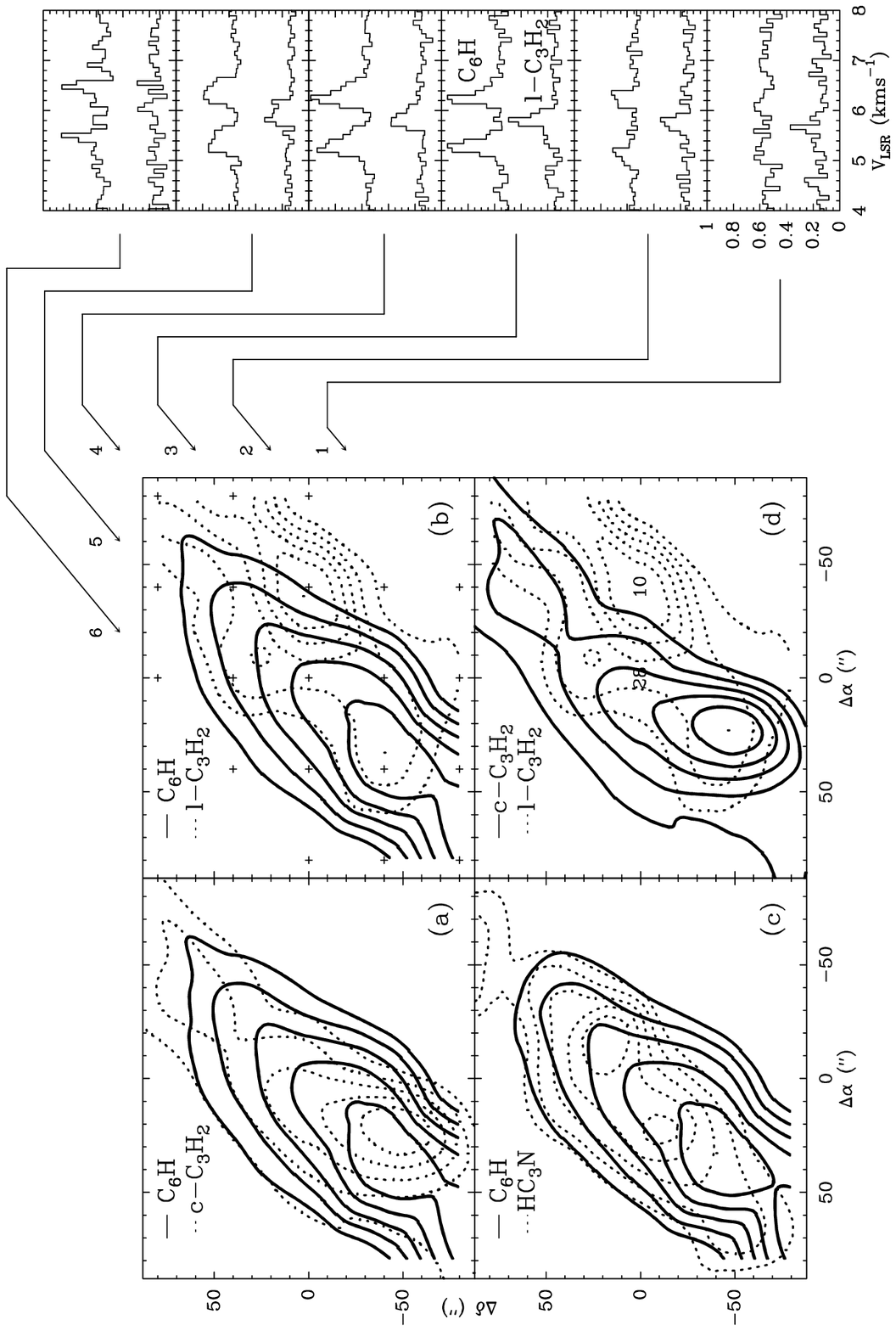]
{Contour levels of integrated intensity in TMC-1 for: {\bf Panel (a):} C$_6$H
$^2{\Pi}_{3/2}$ J=15/2--13/2 and
{\cCCCHH} $\rm 1_{10}-1_{01}$; {\bf Panel (b):} C$_6$H $^2{\Pi}_{3/2}$
J=15/2--13/2 and {\lCCCHH} $\rm 1_{01}-0_{00}$ (the crosses indicate the
measured positions); {\bf Panel (c):} C$_6$H
$^2{\Pi}_{3/2}$ J=15/2--13/2 and
HC$_3$N 10--9; {\bf Panel (d):} {\cCCCHH} $\rm 1_{10}-1_{01}$ and {\lCCCHH}
$\rm 1_{01}-0_{00}$ (numbers indicate the values of R$_2$, i.e. the ratio
{\it N}(\cCCCHH)/{\it N}(\lCCCHH)).
All contours go from 50 to 100 \% of the peak value by steps of 10 \%. The peak
values are 0.16 K$\cdot$kms$^{-1}$, 2.6  K$\cdot$kms$^{-1}$,
0.15  K$\cdot$kms$^{-1}$, and 1.75  K$\cdot$kms$^{-1}$ for
C$_6$H, \cCCCHH, \lCCCHH, and HC$_3$N respectively.
The coordinates of the (0,0) position are:
$\alpha$ = 04$^h$38$^{m}$38.6$^s$,
$\delta$~=~25$^{\circ}$35'45.0'' (1950.0) corresponding to the cyanopolyyne peak.
The boxes to the right compare the C$_6$H and {\lCCCHH} spectra
averaged along the six directions indicated by the arrows. Note the
significant emission of
{\lCCCHH} toward the west side of the TMC-1 filament where the
emission of C$_6$H
decreases rapidly.}

\figcaption[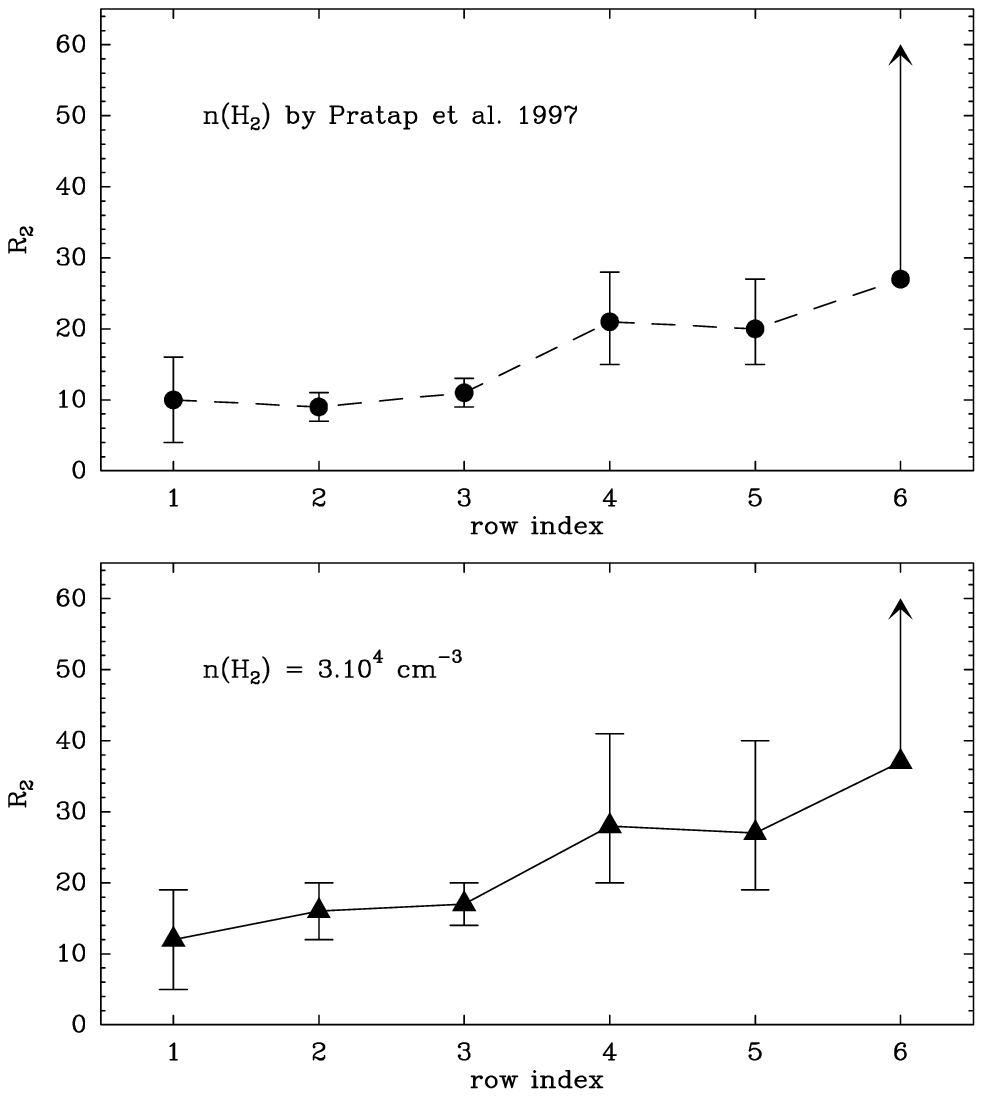]
{Variations of R$_2$ (i.e. the ratio
{\it N}(\cCCCHH)/{\it N}(\lCCCHH))
across the TMC-1 filament using: {\bf Lower panel:} a constant density of 
3$\times$10$^4$ cm$^{-3}$; {\bf Upper panel:} densities derived by Pratap et al.
(1997). In order to increase the signal to noise ratio, the
spectra have been averaged along six rows (separated by 28'') parallel to the
TMC-1 filament. Lower limits in row number 6 are derived from 3$\sigma$ upper 
limits on the \lCCCHH\, emission.
The cyanopolyyne peak (CP) is located within row number 4.}

\figcaption[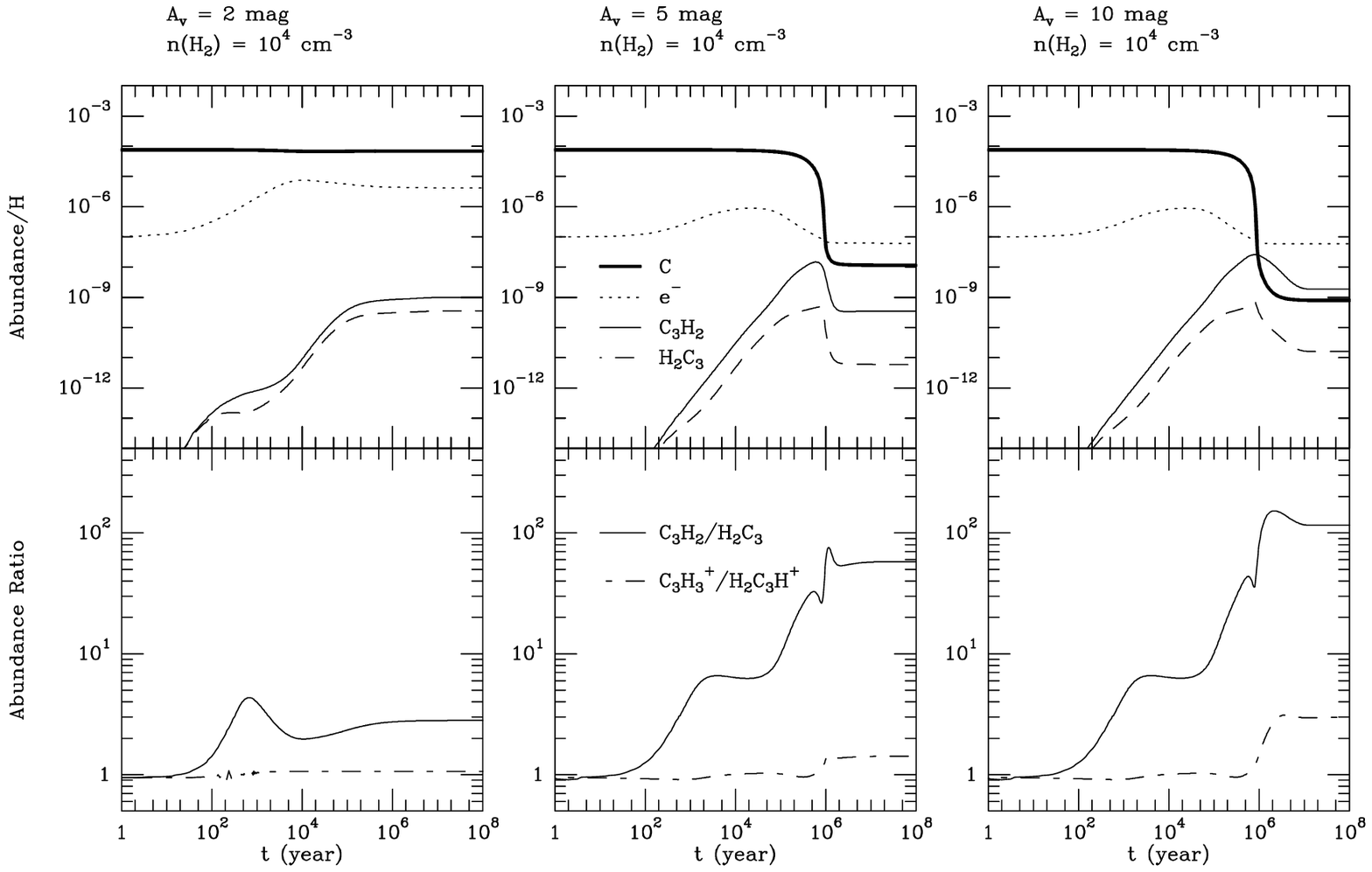]
{Chemical evolution for three different visual extinctions at n(H$_2$) = 10$^4$ cm$^{-3}$
 and T = 10 K. Upper boxes show abundances
relative to H for C, \lCCCHH, \cCCCHH\, and e$^-$. Lower boxes show the resulting cyclic
over linear abundance ratio for C$_3$H$_2$ and C$_3$H$_3^+$ isomers.}

\figcaption[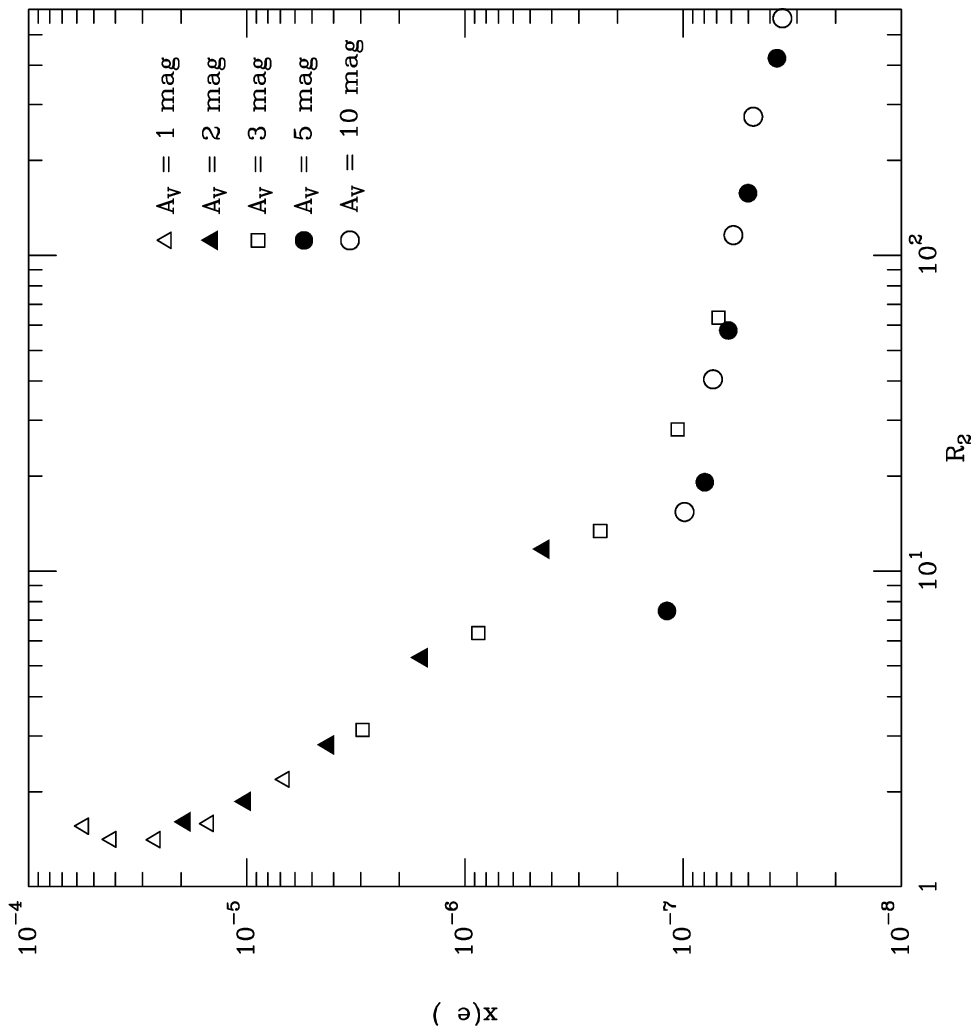]
{The predicted electronic abundance vs. R$_2$ at steady state.
Two regimes can be distinguished according as the visual extinction is low or high.}

\figcaption[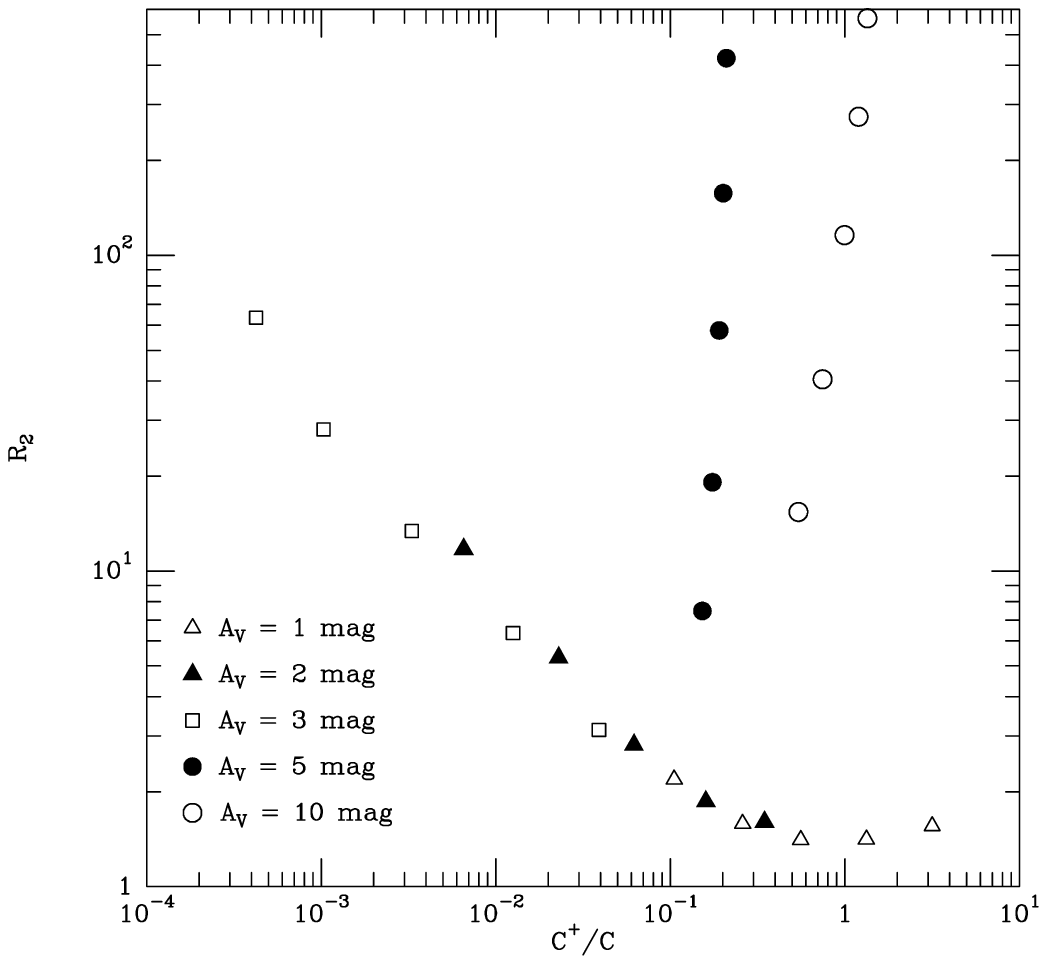]
{The predicted dependence of R$_2$ with the C$^+$/C ratio at steady state. A
clear relationship is established at low visual extinction between R$_2$ and the 
ionization fraction of atomic carbon. Note that at high visual extinctions R$_2$ 
depends
only weakly on C$^+$/C.}

\figcaption[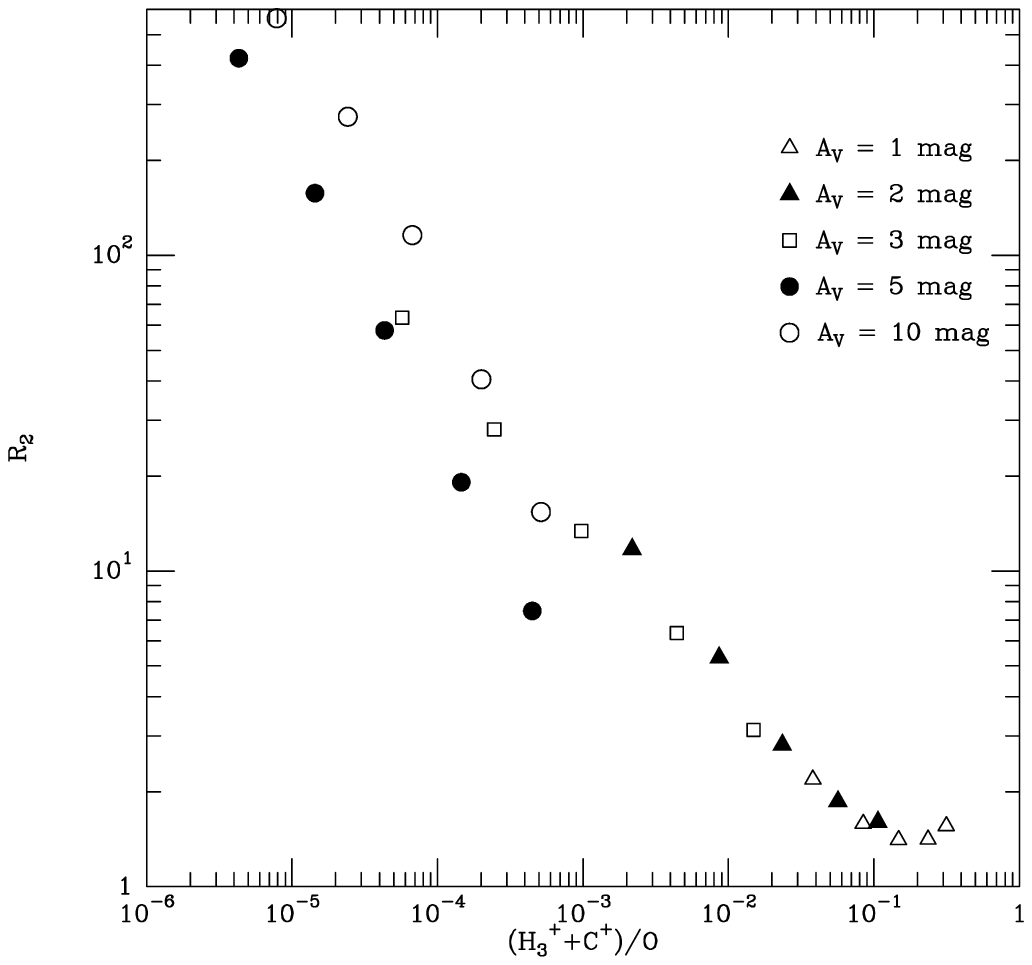]
{Model prediction for the relation of R$_2$ vs. the (H$_3^+$+C$^+$)/O ratio 
at steady state.
Note the good correlation between R$_2$ and 
(H$_3^+$+C$^+$)/O at high visual extinctions.}

\clearpage
\begin{figure}[h]
\plotone{figure1_apj.ps}
\end{figure}

\clearpage
\begin{figure}[h]
\plotone{figure2_apj.ps}
\end{figure}

\clearpage
\begin{figure}[h]
\plotone{figure3_apj.ps}
\end{figure}

\clearpage
\begin{figure}[h]
\plotone{figure4_apj.ps}
\end{figure}

\clearpage
\begin{figure}[h]
\plotone{figure5_apj.ps}
\end{figure}

\clearpage
\begin{figure}[h]
\plotone{figure6_apj.ps}
\end{figure}

\clearpage
\begin{figure}[h]
\plotone{figure7_apj.ps}
\end{figure}

\clearpage

\end{document}